\documentstyle[12pt,epsfig]{article}
\begin{document}
\begin{center}
{\large \bf { BACKWARD ELASTIC $p\ ^3He$
SCATTERING AT  ENERGIES 1 - 2 GeV}}
\vskip 1.5 em {\normalsize {L.D. Blokhintsev,\footnote {Institute of
Nuclear
Physics{\underline ,} Moscow State University, 119899 Moscow ,Russia }
 A.V.Lado,\footnote {Department of Physics, Kazakh State University,
Timiryazev Str., 47, 480121 Alma-Ata, Kazakhstan }
Yu.N. Uzikov \footnote {{\it e-mail address}:
uzikov@nusun.jinr.dubna.su
FAX: (7) 09621 66666}}}

{\it Laboratory of Nuclear Problems, Joint Institute for Nuclear
Research,
 Dubna, Moscow reg., 141980 Russia}
\end{center}
\baselineskip =1.5pc
\vskip 1em
\section*{Abstract}
The two-body transfer amplitude  for  the rearrangement process
 $i+\{jkl\}\to j+\{ikl\}$ is constructed  on the basis of technique
 of 4-dimensional covariant nonrelativistic graphs. The developed
formalism
 is applied to describing  backward elastic $p^3He$ scattering in the
 energy range $0.5\div 1.7$ GeV.  Numerical calculations
 performed  using the 5- channel wave function
 of the $^3He$ nucleus show that the transfer of a noninteracting np-
pair
 dominates and explains satisfactorily the energy  and angular
 dependence of the differential cross section at energies $0.9\div
1.7$ GeV.
 A weak sensitivity to high momentum components
 of  the $~^3He$ wave function in spite of large momentum
transfer  as
 well as a
 very important role of rescatterings in the initial and final states
are
 established.
\baselineskip =1.5pc

 PACS 21.45.+v,  25.40.Cm, 25.55.-e

 KEY WORDS: Faddeev equation, two-nucleon transfer, structure of
$^3He$
\newpage
 The most interesting  peculiarity of the
backward elastic $p\ ^3He$ scattering  at energies  $0.5\div 1.5$ GeV
 is the large momentum transfer $\Delta = 2\div 3$ GeV/c
\cite{berth81}.
 This value of $\Delta $ is considerably larger than the  $\Delta $
value
reached in the experiments with electron scattering from the
$\ ^3He$ nucleus
and therefore one can hope   to obtain an important new
information about the
structure of $\ ^3He$ at short distances between nucleons
$r_{NN}\sim 1/\Delta $ from the $p\ ^3He $ scattering experimental
data.
Simple models of  the $p~^3He\to ~^3Hep$ process like deuteron
($d$) and
singlet deuteron ($d^*$) exchanges
 suggested previously in Refs.
\cite {barry}- \cite{kobush}
are in agreement with  this point of view and lead  to the
conclusion that
the differential cross section of backward elastic $p^3He$-scattering
at initial energies 1-2 GeV \cite{berth81} "measures"  the high
momentum
components of $^3He$ wave function (w.f.)
in the configurations  $~^3He\to d+p, \, ~^3He\to d^*+p$.
According to Ref. \cite{kobush}, the  momentum
distributions
in these
channels, obtained from the experimental data on
inclusive reactions
 $~^3He\, A\to p(0^o)X$ and $~^3He\, A\to d(0^o)X$,
 are  quite compatible with those from  the $p~^3He\to ~^3Hep$
process.
 However, the main drawback of the papers \cite {barry} -
\cite{kobush}
  is the use of the
 two-body approximation for  the $\ ^3He$ nucleus structure and
the neglect
 of initial and final state interactions in \cite{kobush}.
 As shown here (see \cite{blu}),
 the relation between  the full
 three-body wave function  of  the $^3He$ nucleus and  the cross
section
of  the $p~^3He\to ~^3Hep$ process is nontrivial mainly due to
three-body structure of $^3He$
 and Glauber rescatterings in the initial and final states. We found
that
the deuteron  and singlet deuteron  exchanges are negligible as
 compared to  the noninteracting np-pair exchange which allows
to  explain the
available experimental data at $T_p\sim 0.9 -1.7 GeV$  without
involving high momentum components of the $^3He$ w.f.

We proceed from the technique of covariant 4-dimensional
(nonrelativistic)
Feynman graphs \cite {blokh67}.
The dominating mechanism displays weak sensitivity to the relativistic
 effects. For this reason the use of the nonrelativistic technique is
valid.
When processes involving three-body bound states  $\{ijk\}$ are
considered{\underline ,}
the main element of this  approach is the amplitude of the virtual
 decay of such a
 state into three particles $\{ijk\}\to i+j+k$. The simplest
analytical
behaviour is manifested by  the truncated part of this amplitude,
$R^{ij}({\bf q}_{ij},
{\bf Q}_k,E_{ij}$), which is the sum of those graphs  which end in the
interaction
between particles only with indices $i$ and $j$. Here ${\bf q}_{ij}$
and
 $E_{ij}$ are the momentum and energy of relative motion in the $ij$
pair
 respectively, ${\bf Q}_k$ is the momentum of the nucleon spectator
$k$
in the center-of-mass system (c.m.s.) of $\{ijk\}$ nucleus.
The properties of vertex functions $R^{ij}$ and corresponding integral
equations are present{\underline ed} in Ref. \cite{blokh67}. The most
general formula for the
two-nucleon transfer amplitude in the process
4 + \{123\}$\to $ 1 + \{423\}
 is given by the  sum of graphs in Fig.1.
The graph $a$ in Fig.1 describes the  transfer
 of two noninteracting nucleons with numbers 2 and 3. The second
graph $({\it b})$ in Fig.1 describes the interacting pair transfer
(IPT),
which contains the deuteron  and singlet deuteron  exchanges.
After performing the integration over  relative energy of  the
transferred pair $E_{23}${\underline ,} the first term
transforms to the infinite series of 3-dimensional
graphs. In the Born approximation the sum of graphs in Fig.1 can be
written
as
$$T_B=6(2\pi)^{-3}\int d^{3}{ q}_{23}\,
 (\varepsilon +{\bf q}^2_{23}/m+3\,{\bf Q}^2_1/4m)\chi_
{p'}^+(1)\ \{\varphi_f^{23^+}(4;23)\varphi_i^{31}(2;31)$$
\begin{equation}\label{helio1}+\varphi_f^{42^+}(3;42)\varphi_i^{31}(2;
+\varphi_f^{34^+}(2;34)\varphi_i^{31}(2;31)\}\chi_p(4),
\end{equation}
where  $\varphi^{ij}(k;ij)=\varphi^{ij}({\bf q}_{ij},{\bf Q}_k)$ is
the Faddeev component of the full $^3He$ wave function \cite{faddeev}
$\Psi_{\{123\}}=\varphi^{12}+\varphi^{23}+\varphi^{31}$ (the
 lower indices
$i$ and $f$  denote the initial and final states respectively);
$\chi_p(1)$ and $\chi_p(4)$ are the spin-isospin wave functions of
 the protons;
$\varepsilon$ is the binding energy of $^3He$ nucleus and $m$ denotes
the nucleon mass.
The terms $\varphi^{23^+}\varphi^{31},\ \varphi^{42^+}\varphi^{31}, \
\varphi^{34^+}\varphi^{31}$ correspond to the IPT, sequential
transfer (ST)
in the Born approximation  and nonsequential transfer (NST) amplitudes
respectively .

 The most important feature  of  the integral (\ref{helio1})  is
 the following.
The term $\varphi_f^{42^+}\varphi_i^{31}$ in the integrand of Eq.
 (\ref{helio1}), which is responsible for the dominating
ST
 contribution, has the following structure of arguments
$$\varphi_f^{42^+}\varphi_i^{31}=\varphi_f^{42^+}(-{1\over2}{\bf
q}_{23}-
{3\over4}{\bf Q}_4,{\bf q}_{23}-{1\over2}{\bf Q}_4)$$
\begin {equation}\label {p3h-85}\times\
\varphi_i^{31}(-{1\over2}{\bf q}_{23}+{3\over4}{\bf Q}_1,-{\bf
q}_{23}-
{1\over2}{\bf Q}_1).
\end{equation}
The remaining two terms corresponding to the NST and IPT mechanisms
have
the  forms
$$\varphi_f^{34^+}\varphi_i^{31}=\varphi_f^{34^+}(-{1\over2}{\bf
q}_{23}+
{3\over4}{\bf Q}_4,-{\bf q}_{23}-{1\over2}{\bf Q}_4)$$
\begin {equation}
\label {p3h-86}
\times
\varphi_i^{31}(-{1\over2}{\bf q}_{23}+{3\over4}{\bf Q}_1,-{\bf
q}_{23}-
{1\over2}{\bf Q}_1),
\end{equation}
\begin {equation}\label {p3h-87}
\varphi_f^{23^+}\varphi_i^{31}=\varphi_f^{23^+}({\bf q}_{23},{\bf
Q}_4)
\varphi_i^{31}(-{1\over2}{\bf q}_{23}+{3\over4}{\bf Q}_1,-{\bf
q}_{23}-
{1\over2}{\bf Q}_1).
\end{equation}
At the scattering  angle $\theta_{c.m.}=180^\circ$  we get
 ${\bf Q}_4=-{\bf Q}_1$.
Under this condition Eq. (\ref{p3h-85}) differs from Eqs.
 (\ref{p3h-86}) and (\ref{p3h-87}) in that two of four momenta can
 simultaneously  became equal to zero at integration over
 ${\bf q}_{23}$ (for example, at the point
 ${\bf q}_{23}=3/2\ {\bf Q}_1$ one has  $-1/2\ {\bf q}_{23}-3/4\
{\bf Q}_4=0$ and $-1/2\ {\bf q}_{23}+3/4\ {\bf Q}_1=0)$. In Eqs.
 (\ref{p3h-86}) and (\ref{p3h-87}) only one momentum can be equal to
zero
while the other three have the values
 $\sim \vert{\bf Q}_1\vert=\vert{\bf Q}_4\vert$.
In the kinematic region $T_p=0.7-1.7\ GeV$
the momenta $\vert{\bf Q}_1\vert$ and  $\vert{\bf Q}_4\vert$ have
rather
large values $\sim0.5\ GeV/c$.
  One can conclude from this that term (\ref{p3h-85})   makes
 the dominating contribution to  the integral  (\ref{helio1})
since the functions $\varphi^{ij}({\bf q},{\bf Q})$ decrease
fast with increasing $\vert{\bf q}\vert$ or $\vert{\bf Q}\vert$.
Numerical calculations are performed using  Faddeev $^3He$
w.f.
from \cite{kim,hgs} for  NN-potential in $^1S_0$ and $^3S_1-^3D_1$
states
in the RSC form.
This wave function consists of 5  channels for each Faddeev
component
$\varphi ^{ij}$.
As shown in Fig.2,  in the region $T_p=1-1.7 GeV$ the ST contribution
is by factor $\sim 30-40$ higher in comparison  with the
experimental data
and by 3-4 orders of magnitude larger than the IPT and NST-
contributions.
 At $\theta_{c.m}< 180^o$ the ST amplitude loses this advantages and
as
 a result its contribution decreases very fast with increasing the
difference
$|180^o-\theta_{c.m.}|$,  whereas  the experimental angular dependence
of the cross section is smooth (Fig.3).

  Taking into account the rescatterings in the initial and final
states
of the process $p^3He\to ^3Hep$  in the framework of diffraction
multistep
theory of Glauber-Sitenko \cite{sitenko70}, we find
for the $np$ transfer amplitude the following expression
$$T_{fi}^{dist}=T_B({\bf p}'_\tau,{\bf p}'_p;{\bf p}_\tau,{\bf
p}_p)+{i\over4\pi
p_p}\int d^2q\ F_{p\tau}({\bf q})T_B({\bf p}'_\tau,{\bf p}'_p;{\bf
p}_\tau+{\bf q},
{\bf p}_p-{\bf q})$$
 $$+{i\over4\pi p_{p'}}\int d^2q'\ f_{pp}({\bf q}')T_B({\bf p}_
\tau'-{\bf q}',{\bf p}'_p+{\bf q}';{\bf p}_\tau,{\bf p}_p)$$
\begin{equation}
\label{helio2}
-{1\over(4\pi)^2p_{p'}\,p_p}\int\int d^2qd^2q'\ F_{p\tau}({\bf
q})f_{pp}({\bf q}')
T_B({\bf p}_\tau-{\bf q}',
{\bf p}'_p+{\bf q}';{\bf p}_\tau+{\bf q},{\bf p}_p-{\bf q});
\end{equation}
here the amplitude $T_B$ is defined by Eq. (\ref{helio1}) and
the amplitudes $F_{p\tau}({\bf q})$ and $f_{pp}({\bf q})$ describe the
forward elastic $p^3He$  and $pp$-scattering,
 respectively, at the transferred momentum {\bf q};
${\bf p}_\tau$ and ${\bf p}_p\, ({\bf p}'_\tau$ and ${\bf p}'_p)$
 are the c.m.s momenta of initial
 (final)  $^3He$ nucleus and proton, respectively.

On the basis of Eq. (\ref{helio2}) we found that (i) the absolute
value
of the differential cross section at $\theta_{c.m.}=180^o$ decreases
considerably  due to rescatterings and
as a result agrees satisfactorily  with the experimental data at
$T_p=0.7-1.7 GeV$ \cite{berth81} (Fig.2);
 (ii) the angular dependence becomes very smooth in the
range $\theta_{c.m.}=160^0-180^0$ in qualitative agreement with the
experimental data (Fig.3).
 The  cross section at $\theta_{c.m.}=180^o$ decreases due to
the
imaginary part of  the elastic $pN$- scattering amplitude, which
takes into account  the  coupling to inelastic channels.
 As shown recently in Ref. \cite{uz97},
the suppression factor caused by  rescatterings
in the  framework of  the same method is about $\sim 2-3$
for the  one-nucleon exchange mechanism of  the process
$pd\to dp$. The smooth angular dependence comes from the fact
 that due to  rescatterings the effective  arguments of
 the w.f. of $^3He$ in the integrals (\ref{helio2})
decrease in the upper and the  lower vertices
simultaneously \cite{uz97}. Weak sensitivity of the dominating ST-
mechanism
to the high momentum components of $^3He$ w.f. is manifested by (i)
the nonimportant
contribution of D-components of w.f.  at $T_p>1 GeV$ (see Fig.2) and
 (ii)  very small enhancement of the cross section after
 substitution
of Lorenz-invariant momenta ${\bf q}_{ij}\, {\bf Q}_k$  into the
amplitude
$T_B$  instead of nonrelativistic ones. From the other hand, both
the D-component of  the $^3He$ w.f. and relativistic effects
 are very important for the mechanism
of deuteron exchange. However the role of that mechanism is
negligible.

In conclusion, an unexpected feature of process $p^3He\to ^3Hep$
at $T_p=1-1.7 GeV$ is established, namely,
a weak sensitivity of the dominating ST mechanism to high momentum
 components
 of $~^3He$ wave function in spite of large momentum transfer.
It should be noted  that the  ST amplitude reduces exactly to
zero if
the component ${\hat A}\{ NN(^1S_0)+N\}$, where $\hat A$ is the
antisymmetriser, is excluded from the $^3He$ w.f. According to
the
definition in work \cite{hgs}, it corresponds to
the channel with number $\nu=1$.
Consequently the cross section of  the process $p^3He\to ^3Hep$
at
$\theta_{c.m.}=180^o$ measures the weight of this channel
\cite{luz93}.
Faddeev calculations \cite{sauer} predict that
the ground state spin of  the $^3He$
nucleus is dominated by the neutron just due to this component of
 the $^3He$ w.f. Therefore it is of great interest to perform
 the
experimental investigation of spin observables in this process at
 energies $T_p>1 GeV.$

 This work  was supported in part by  the Russian
 Foundation
 for Basic Researches (grant $N^o$ 96-02-17458).

\eject

\eject
\section*{Figure captions}

Fig.1.  Two-particle transfer amplitude of the
4 + \{123\}$\to $ 1 + \{423\} process in terms of 4-dimensional
graphs:
{\it a} -- the noninteracting pair transfer,

{\it b} -- the interacting pair transfer (IPT),

{\it c} -- total amplitude; $R_i'=R^{12}_i+R^{31}_i,\
 R_f'=R^{42}_f+R^{34}_f, \,
R_i=R^{12}_i+R^{31}_i+R^{23}_i$

Fig.2. The differential cross section of the elastic $p^3He$
scattering
in the c.m.s. at $\theta_{c.m.}=180^o$ as a function of initial
proton energy
$T_p$ and transferred momentum $\Delta $. The curves are the results
of
calculations ( curves 1 -- 5 correspond to the Born
approximation):
 1 - IPT (S) with $S$ component of  the \ $^3He$ w.f.,
2 -- NST (S), 3 -- ST (S), 4 -- IPT (S+D), 5 -- ST (S+D);
6 -- ST (S) taking into account rescatterings.
The dots   denote the data from Ref.\cite {berth81}.

Fig.3. The differential cross section of the $p^3He$ elastic
scattering
into backward hemisphere at different initial energies (MeV) shown
near
the curves.The curves are the results of calculations using
IPT+NST+BST mechanism in the S-wave approximation for the $^3He$
wave function : the dashed curves are obtained in  the Born
approximation  and the full curves are obtained taking into account
Glauber rescatterings; $p$ is the c.m. momentum of  the proton,
$\theta_{c.m.}$ is the
scattering angle. The experimental dots are from Ref. \cite {berth81}.

\newpage
\begin{figure}[h]
\mbox{\epsfig{figure=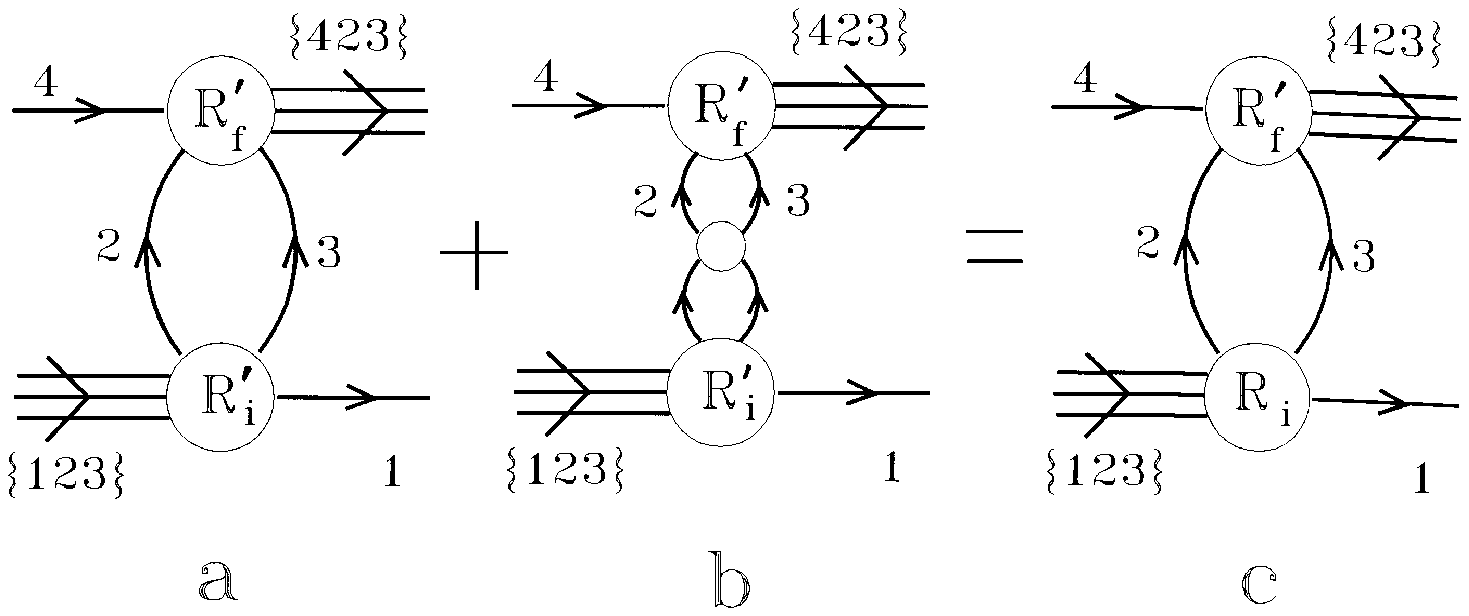,height=0.25\textheight, clip=}}
\caption{}
\end{figure}
\newpage
\begin{figure}[h]
\mbox{\epsfig{figure=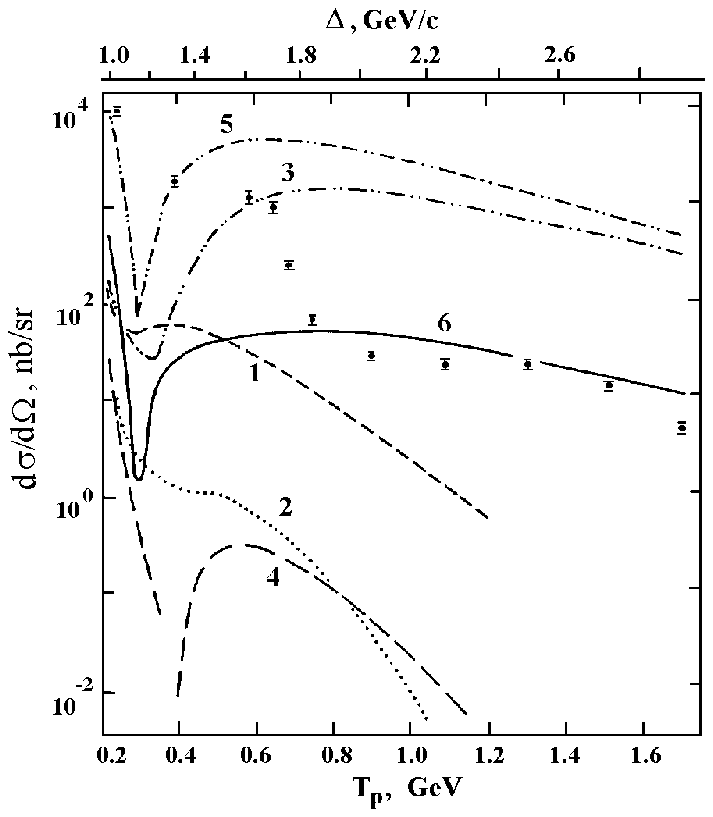,height=0.7\textheight, clip=}}
\caption{}
\end{figure}

\newpage
\begin{figure}[h]
\mbox{\epsfig{figure=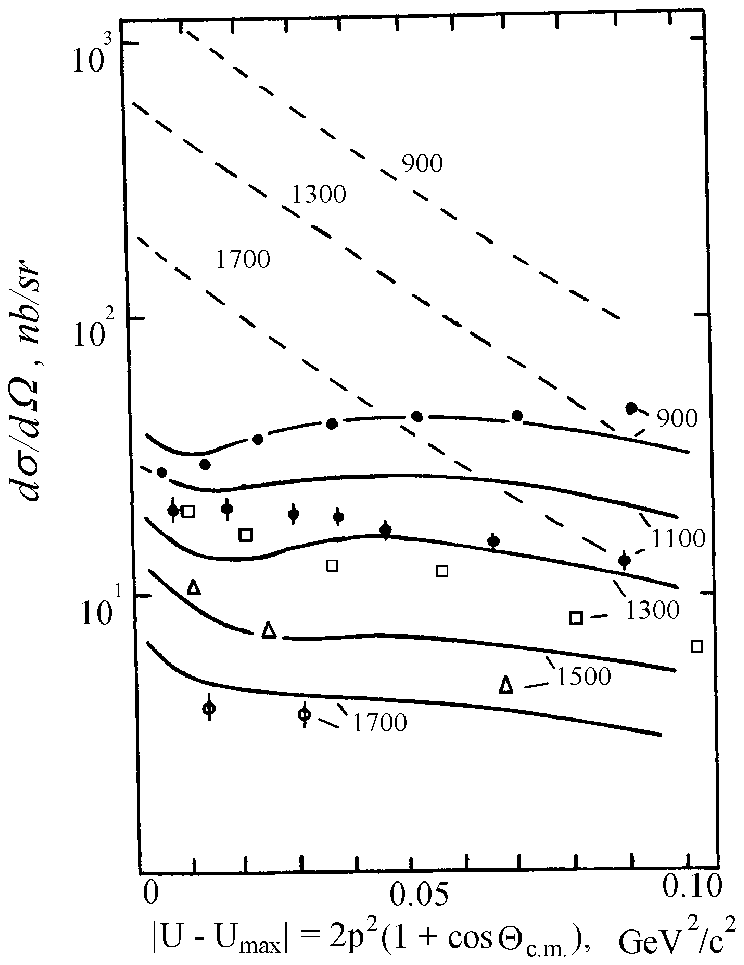,height=0.70\textheight, clip=}}
\caption{}
\end{figure}

\end{document}